\documentclass[10pt,letterpaper]{article}
\usepackage[top=0.85in,left=2.75in,footskip=0.75in]{geometry}

\usepackage{amsmath,amssymb}

\usepackage{changepage}

\usepackage{textcomp,marvosym}

\usepackage{cite}

\usepackage{nameref,hyperref}

\usepackage[nopatch=eqnum]{microtype}
\DisableLigatures[f]{encoding = *, family = * }

\usepackage{subcaption}% Min: Added temporarily for old the-anh figure MARK
\usepackage{listings}% Min: Added for commands-and-versions section

\usepackage[table]{xcolor}

\usepackage{array}

\newcolumntype{+}{!{\vrule width 2pt}}

\newlength\savedwidth

\raggedright
\usepackage[aboveskip=1pt,labelfont=bf,labelsep=period,justification=raggedright,singlelinecheck=off]{caption}

\makeatletter
\renewcommand{\@biblabel}[1]{\quad#1.}
\makeatother

\usepackage{lastpage,fancyhdr,graphicx}
\usepackage{epstopdf}
\fancyheadoffset[L]{2.25in}
\fancyfootoffset[L]{2.25in}
\begin{document}
% Min: begin MARK added for commands-and-versions
\definecolor{codegreen}{rgb}{0,0.6,0}
\definecolor{codegray}{rgb}{0.5,0.5,0.5}
\definecolor{codepurple}{rgb}{0.58,0,0.82}
\definecolor{backcolour}{rgb}{0.95,0.95,0.92}
\lstdefinestyle{mystyle}{
    backgroundcolor=\color{backcolour},   
    commentstyle=\color{codegreen},
    keywordstyle=\color{magenta},
    numberstyle=\tiny\color{codegray},
    stringstyle=\color{codepurple},
    basicstyle=\ttfamily\footnotesize,
    breakatwhitespace=false,         
    breaklines=true,                 
    captionpos=b,                    
    keepspaces=true,                 
    numbers=left,                    
    numbersep=5pt,                  
    showspaces=false,                
    showstringspaces=false,
    showtabs=false,                  
    tabsize=2
}
\lstset{style=mystyle}
% Min: end MARK added for commands-and-versions

\vspace*{0.2in}

% Title must be 250 characters or less.
\begin{flushleft}
{\Large
\textbf\newline{Improved Community Detection using Stochastic Block Models} % Please use "sentence case" for title and headings (capitalize only the first word in a title (or heading), the first word in a subtitle (or subheading), and any proper nouns).
}
\newline
Minhyuk Park\textsuperscript{1*},
Daniel Wang Feng\textsuperscript{1},
Siya Digra\textsuperscript{1},
The-Anh Vu-Le\textsuperscript{1},
Lahari Anne\textsuperscript{1},
George Chacko\textsuperscript{1*},
Tandy Warnow\textsuperscript{1*}
\\
\bigskip
\textbf{1} Siebel School of Computing and Data Science, University of Illinois Urbana-Champaign, Urbana, Illinois, United States of America
\\
\bigskip

% Insert additional author notes using the symbols described below. Insert symbol callouts after author names as necessary.
% 
% Remove or comment out the author notes below if they aren't used.
%
% Primary Equal Contribution Note
% \Yinyang These authors contributed equally to this work.

% Additional Equal Contribution Note
% Also use this double-dagger symbol for special authorship notes, such as senior authorship.
% \ddag These authors also contributed equally to this work.

% Current address notes
% \textcurrency Current Address: Dept/Program/Center, Institution Name, City, State, Country % change symbol to "\textcurrency a" if more than one current address note
% \textcurrency b Insert second current address 
% \textcurrency c Insert third current address

% Deceased author note
% \dag Deceased

% Group/Consortium Author Note
% \textpilcrow Membership list can be found in the Acknowledgments section.

% Use the asterisk to denote corresponding authorship and provide email address in note below.
* \{minhyuk2,warnow,chackoge\}@illinois.edu

\end{flushleft}
% Notes:
% introduction, results and discussion, summary, conclusion, materials and methods
% no supplementary materials if possible
% tiff or eps but tiff preferred and use PACE (also name the files "fig 1")

% Please keep the abstract below 300 words
\section*{Abstract} \label{sec:abstract}
Identifying edge-dense communities that are also well-connected is an important aspect of understanding community structure. Prior work has shown that community detection methods can produce poorly connected communities, and some can even produce internally disconnected communities. In this study we evaluate
the connectivity of communities obtained using Stochastic Block Models. We find that SBMs produce internally disconnected communities from real-world networks.
We present a simple technique, Well-Connected Clusters (WCC), which repeatedly removes small edge cuts until the communities meet a user-specified threshold for
well-connectivity.  Our study using a large collection of synthetic networks based on clustered real-world networks  shows that using WCC as a post-processing tool
with SBM community detection typically improves clustering accuracy. WCC is fast enough to use on networks with millions of nodes and is freely available in open
source form.

% Please keep the Author Summary between 150 and 200 words
% Use first person. PLOS ONE authors please skip this step. 
% Author Summary not valid for PLOS ONE submissions.   
\section*{Author summary}
Using a corpus of both real-world and synthetic networks, we establish that communities found using Stochastic Block Models are often internally disconnected, meaning that the community has two or more components.
We propose a simple technique, Well-Connected Clusters (WCC), for postprocessing the communities found by an SBM by breaking up each community  into smaller communities that are well-connected, according to a user-provided threshold. 
We show that  the WCC postprocessing treatment improves the accuracy of SBM clusterings on a large set of synthetic networks. 
Finally, WCC is very fast and can be used on large networks in the order of ten million nodes. %result in improved clustering accuracies. % 127 words

% \linenumbers

% Use "Eq" instead of "Equation" for equation citations.
\section*{Introduction} \label{sec:introduction}
Community detection, also known as graph clustering, is the problem of taking a graph and partitioning the vertices into disjoint subsets so that each set has some desirable properties, such as edge-density and separability from the rest of the graph.
Another desirable property is well-connectedness which means that each community should not only be internally connected but should not be able to be split into two communities by the deletion of a small number of edges \cite{kannan2004clusterings,traag2019louvain}, i.e., be ``well-connected".

Not much is yet known about the properties of clustering methods with respect to cluster  edge-connectivity.
However, it is known that the Louvain algorithm \cite{blondel2008fast} can produce disconnected communities \cite{traag2019louvain}.
In a recent study published in Complex Networks and Applications (CNA) 2024 \cite{park2024improved-arxiv}, we showed that community detection using Stochastic Block Models  implemented in graph-tool \cite{graph-tool}  produced disconnected communities, and that this tendency increased with the network size so that most communities found by SBM on large networks were disconnected.

Park {\em et al.} \cite{park2024} evaluated clustering methods with respect to the size of the minimum edge cut for each cluster.
Using a mild standard for ``well-connected", by which a cluster of $n$ nodes is considered well-connected only when its minimum edge cut size  is greater than $\log_{10}(n)$ and otherwise is poorly connected, \cite{park2024} found that
 the Leiden algorithm \cite{leiden-code,traag2019louvain}, Infomap \cite{rosvall2009map}, Iterative-K-core Clustering (IKC) \cite{Wedell2022} and Markov Clustering (MCL) \cite{VanDongen2008}
produce poorly connected clusters.

The main goal of our CNA 2024 study \cite{park2024improved-arxiv} was to improve the edge connectivity of SBM clusterings of input networks, produced using  graph-tool \cite{graph-tool}.

We evaluated the benefit of using three post-processing techniques that take an SBM clustering and modify it by splitting clusters into smaller clusters, with the goal of improving the edge-connectivity so that they are well-connected, according to the same mild standard of $\log_{10}n$ used in Park {\em et al.} \cite{park2024} and discussed in the previous paragraph.
We explored the use of the Connectivity Modifier \cite{park2024}, a technique that iteratively finds and removes small edge cuts and then re-clusters until all clusters are well-connected.
In addition, we  presented and studied two simpler techniques:   Well-Connected Clusters (WCC), which iteratively finds and removes small edge cuts but does not re-cluster, and Connected Components (CC), which just returns the components of each cluster as the new communities.

Our CNA 2024 study \cite{park2024improved-arxiv} explored the impact of these post-processing methods on clustering accuracy, using synthetic networks produced by the LFR \cite{lfr-generation-code} method, and established  that each technique was generally helpful in terms of the clustering accuracy measured using the  Adjusted Rand Index (ARI) score \cite{hubert1985comparing}, compared to the untreated SBM clustering. 
However, we found  that using WCC provided the most reliable improvement in accuracy compared to both CM and CC.

In this extended study, we expand on \cite{park2024improved-arxiv} in multiple ways.  
First, we show nearly all clusters in SBM clusterings of bipartite  networks are disconnected, based on which we then restricted the remainder of the study to non-bipartite networks.  Second, we explore SBM clusterings produced using graph-tool (whether followed by one of the post-processing treatments or not) on additional synthetic networks, including  synthetic networks generated using the recently described RECCS method\cite{anne2024synthetic}. We also compare these SBM-based clusterings to the Leiden algorithm \cite{leiden-code} optimizing modularity or the Constant Potts model \cite{traag2019louvain}.
While \cite{park2024improved-arxiv}   explored accuracy only using the Adjusted Rand Index (ARI), here we use three additional criteria.   Finally, we evaluate the computational performance of the treatments for SBM   on large real-world networks.

\section*{Materials and methods} \label{sec:methods}
\subsection*{Networks}
\subsubsection*{Real-world networks}
We used a set of  120 real-world networks: 118 from the Netzschleuder network catalogue  \cite{peixoto2020netzschleuder} and  two other networks from \cite{park2024}. The smallest of these networks has 11 nodes and the largest contains 13,989,436 nodes.
Networks were obtained as edgelists without directionality or weights and pre-processed to remove any self-loops or parallel (duplicate) edges when present. 
See Section D in S1 Appendix for the full list of datasets.

\subsubsection*{Synthetic networks}
We used synthetic networks generated by either the LFR \cite{lancichinetti2008benchmark} or RECCS \cite{anne2024synthetic} methods.
These simulators were given parameters derived from clustered real-world networks, as described below.  
The LFR networks are available through the original publication \cite{park2024}, and 
the RECCS networks  are available at \cite{illinoisdatabankIDB-9805305}.

\paragraph{LFR synthetic networks}
The 27 LFR networks in this study are from a prior study \cite{park2024} and are already in the public domain.  
Each of these networks was produced using the LFR software, using parameters obtained from a clustered real-world network.
The real-world  networks on which they are based are cit\_hepph, cit\_patents, wiki\_topcats, the Curated Exosome Network (CEN) and Open Citations (OC), and 
the  networks were clustered  using Leiden optimizing either modularity or the Constant Potts Model (CPM) using different resolution values  (0.0001, 0.001, 0.01, 0.1, and 0.5).
As reported in \cite{park2024}, one of these LFR networks failed and two LFR networks produced a large proportion of disconnected ground-truth clusters and were not included in \cite{park2024}.
Thus, each empirical network produced up to six LFR networks.

The smallest of these LFR networks has 34,546 nodes (the networks based on cit\_hepph) and the largest has 3,774,768 nodes (the networks based on cit\_patents). 
Due to scalability issues with the LFR software, the LFR networks based on the two largest real-world networks (CEN and OC) were limited to 3 million nodes.

\paragraph{RECCS synthetic networks}
To produce RECCS networks, we provided parameters to the RECCS software for the Leiden-CPM(0.01) clustering of the real-world networks (i.e., cit\_hepph, cit\_patents, and wiki\_topcats) ranging in size from 34,546 nodes to 3,775,768 nodes. 

\subsection*{Stochastic Block Models} \label{sec:methods-sbm}
For each real-world network, we used  the graph-tool package \cite{graph-tool} to produce a ``selected" SBM clustering, with the following protocol.
First,  we clustered the network using three different models of SBM: degree-corrected \cite{dcsbm}, non degree-corrected \cite{holland1983sbm}, and planted partition \cite{ppsbm}. 
We then selected the clustering that achieved the lowest description length. 
This clustering is referred to as the 
``Selected SBM'' clustering for the real-world network. 

\subsection*{Post-processing treatments to improve connectivity}
We post-processed clusters in three ways, each of which is designed to improve the edge-connectivity of the clusters in a given clustering.

\paragraph{The Connectivity Modifier (CM)}This approach was presented in \cite{park2024}, and 
assumes that the user has specified a minimal requirement for edge-connectivity.
CM takes an input clustering, produced using a specified clustering method, and then repeatedly cuts clusters into smaller clusters as follows.
Given a cluster, if it has an edge cut that is below the minimal required size, then that edge cut is deleted, resulting in splitting the cluster into two parts.
Then each part is re-clustered, using the same method that produced the original clustering.
The newly produced clusters are then subject to the same treatment, until all clusters are considered well-connected, in that they pass the minimal threshold for edge cut size. 

The CM pipeline optionally allows the user to specify a minimum cluster size, so that any cluster below that size is deleted.
In \cite{park2024}, we found that removing small clusters (below size 11) reduced clustering accuracy, and so is not desirable.
In this study, therefore, we run CM without the optional removal of small clusters.
We also use the default setting for well-connectedness of $\log_{10}(n)$, where $n$ is the number of vertices in the cluster. CM is the most complex technique of the three methods. 

\paragraph{Well-Connected Clusters (WCC)}
The WCC technique is a simplification of the CM technique that does not re-cluster during the iterative process.  Thus, given a clustering of a network  and a threshold for well-connectedness, WCC checks each cluster to see if it is well-connected.
If so, it puts the cluster in the output, and otherwise it removes the small edge cut from the cluster, thus producing two smaller clusters. The process iterates until each cluster is well-connected.
As with CM, we do not remove any small clusters, and we use the same default setting  for well-connectedness. Again as with CM, we use VieCut from \cite{henzinger2018practical} to calculate a minimum edge cut for a cluster.

\paragraph{Connected Components (CC)}  Given a cluster that is internally disconnected, we replace it with its connected components. CC is the 
simplest of the three methods.

\subsection*{Evaluation}  
Given a clustering of a real-world network, we report the proportion of clusters that are well-connected, poorly-connected, and disconnected, using $\log_{10}(n)$ for the lower bound on the edge cut size for a cluster to be considered well-connected.
We also report the node coverage, which is the percentage of nodes in clusters of size at least two.

We also use synthetic networks with  ground-truth communities to evaluate clustering accuracy.  
We report the Normalized Mutual Info (NMI) \cite{thomas2006elements}, 
Adjusted Rand Index (ARI) \cite{hubert1985comparing},
Adjusted Graph-aware Rand Index (AGRI) \cite{poulin2020comparing}, and Reduced Mutual Information (RMI) scores \cite{newman2020improved} measures.

\subsection*{Infrastructure}
This work made use of the Illinois Campus Cluster, a computing resource that is operated by the Illinois Campus Cluster Program (ICCP) in conjunction with the National Center for Supercomputing Applications (NCSA) and which is supported by funds from the University of Illinois at Urbana-Champaign. Additionally, this research was supported in part by the Illinois Computes project which is supported by the University of Illinois Urbana-Champaign and the University of Illinois System. Each method was allowed up to 72 hours of runtime, 256GB of RAM, and 16 cores of parallelism. If a clustering method failed to complete on the network, we report this with either ``time-out" or ``out-of-memory", depending on the reason for the failure. 

\subsection*{Experiments}
We conducted four experiments:
\begin{itemize}
\item Experiment 1: We evaluate empirical properties of SBM clusterings on real-world networks.
\item Experiment 2: We evaluate the impact of our three treatments on SBM clusterings on real-world networks.
\item Experiment 3: We compare SBM followed by our three treatments against other clustering methods for clustering accuracy on synthetic networks with ground-truth.
\item Experiment 4: We evaluate the computational performance of SBM and our treatments on large real-world networks.
\end{itemize}

\section*{Results}\label{sec:results}
All SBM clusterings in our study were computed using graph-tool \cite{graph-tool}; hence, when we refer to SBM we specifically mean the graph-tool SBM community detection method.

\subsection*{Experiment 1: Empirical properties of SBM clusterings}

\begin{figure}[!h]
\centering
\includegraphics[]{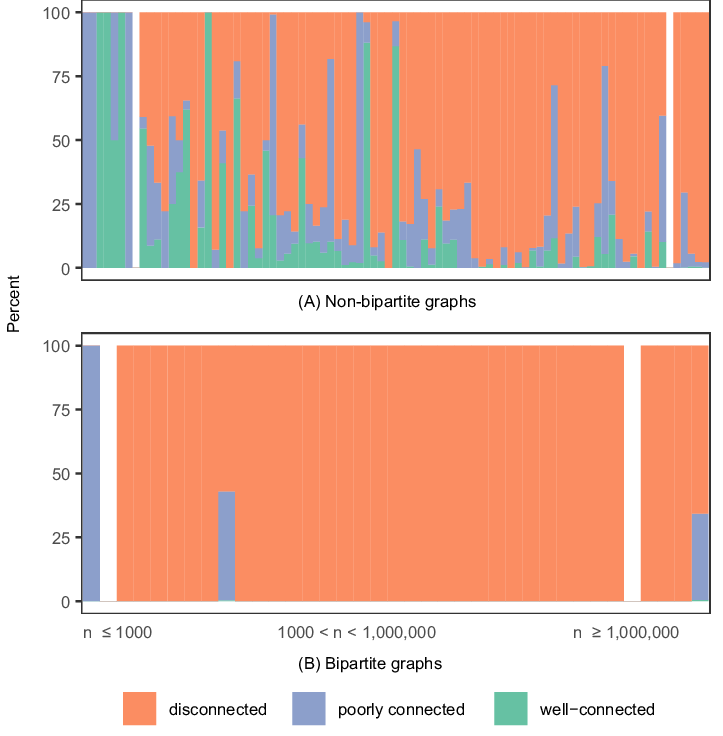}
\caption[Experiment 1: Cluster connectivity of SBM on real-world networks]{\textbf{Experiment 1: Cluster connectivity of SBM on real-world networks.} (A) 85 non-bipartite graphs, (B) 35 bipartite graphs.  SBM often produces poorly-connected and disconnected clusters, with greater tendency on bipartite graphs. The figure shows  the proportion of well-connected (green), poorly connected (blue), and disconnected clusters (orange) in the output clusterings of the lowest description length SBM. The x-axis shows the different real-world networks ordered by number of nodes. White bars separate small, medium, and large networks.}
\label{fig:1}
\end{figure}

For each of 120 networks, ranging in size from 11 nodes to 13,989,436 nodes, we examined the cluster connectivity of the SBM clustering with the lowest description length (selected SBM) (Fig~\ref{fig:1}), 

Of these 120 networks, 35 are bipartite and 85 are non-bipartite, thus producing two groups of interest.
Each group is further divided by size into (i) networks with at most 1000 nodes (``small"), (ii) networks with more than 1000 and less than 1 million nodes (``medium"), and (iii) networks with at least 1 million nodes (``large").

On both bipartite and non-bipartite networks, most of the clusters are internally disconnected (Fig \ref{fig:1}), but the frequency of these internally disconnected clusters is much higher for bipartite networks than non-bipartite networks. 
However, there are no disconnected clusters on the small networks (i.e., at most 1000 nodes), 
whether bipartite or not. Poorly connected clusters, clusters that have edge cuts that are at most $\log_{10}(n)$ in size for a cluster $n$ nodes, are seen, even for small networks.  

Given the high frequency of disconnected clusters, we examined the composition of these clusters using whether the clusters contain internally isolated nodes, which are nodes that do not have any neighbors within their cluster.  Recall that the node coverage measures the percentage of the nodes that are in clusters of size at least two.  
As shown in Table A in S1 Appendix, while node coverage can be less than 100\% for the selected SBM clusterings on the small networks, it is always 100\% on the medium and large networks. 
Hence, any 
node  coverage below 100\% in  a CC-treated SBM for a medium or large network indicates a drop in node coverage entirely due to clusters having these internally isolated nodes.

\begin{figure}[!h]
\includegraphics[]{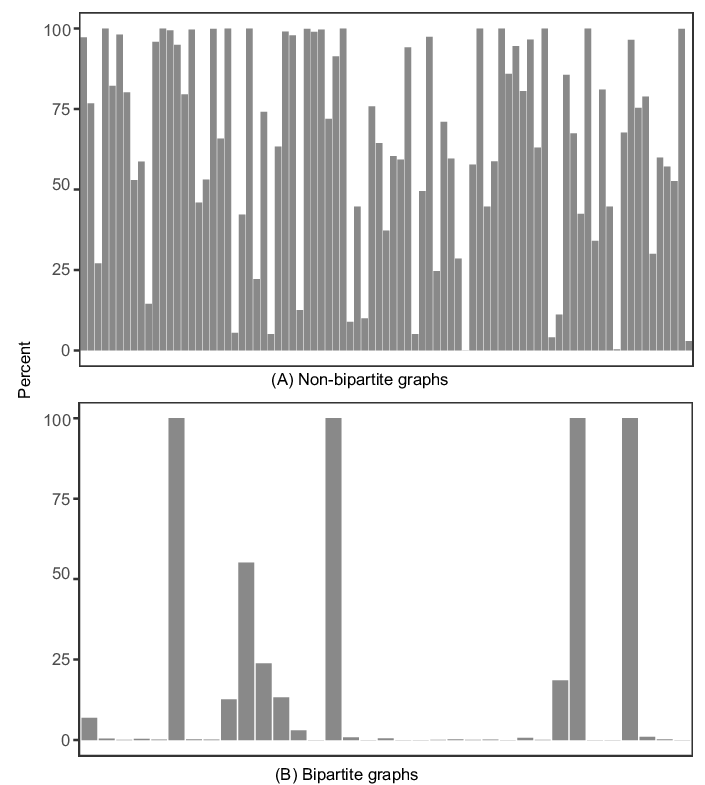}
\caption[Experiment 1: Node coverage of connected components within SBM clusters on non-bipartite and bipartite networks]{\textbf{Experiment 1: Node coverage of SBM+CC clusterings} (A) Node coverage on 85 non-bipartite networks, (B) Node coverage on 35 bipartite networks. The node coverage of SBM treated with CC is much higher on non-bipartite graphs than on bipartite graphs. %%\textcolor{orange}{new figure}
}
\label{fig:2}
\end{figure}

As seen in  Fig~\ref{fig:2}, there is a large drop in node coverage as a result of using the CC treatment for both bipartite and non-bipartite networks, and the drop is very large for bipartite networks.
The reduction in node coverage depends on the network size, with no drop at all for the small non-bipartite networks but a drop from 100\% to 64\% for the medium  non-bipartite networks and from 100\% to 59\% for the large non-bipartite networks (Table A in S1 Appendix).

To summarize, this experiment shows that the selected SBM clusterings tend to have many internally disconnected clusters, and there are many internally isolated nodes in these clusters. 
 The frequency of these internally disconnected clusters increases with the size of the network, but is much higher for bipartite networks than for non-bipartite networks.  
 Because of these observations, we  restricted the rest of the study to the networks that are not bipartite.

\subsection*{Experiment 2: Impact of treatments on real-world networks}
We now consider the impact of CC, WCC, and CM treatments on the selected SBM clusterings on non-bipartite networks.
In general, and as expected, node coverage drops more for CM than for WCC, and more for WCC than for CC (Table A in S1 Appendix).
These trends persist for small, medium, and large real-world networks.

We then examined the distribution of non-singleton cluster sizes of clusterings on the  medium and large networks (Fig A in S1 Appendix).
The median cluster sizes are much higher for the selected SBM than for the post-processed SBMs; this is expected since these treatments break up clusters.
Interestingly, the median cluster sizes for CM-treated clusterings are larger than for CC- or WCC-treated clusterings, and the largest clusters for CM-treated clusterings  are smaller than the largest clusters for the CC and WCC clusterings.

We also see differences in number of non-singleton clusters (Table C in S1 Appendix): the CM treatment produces the fewest non-singleton clusters across all dataset groups, but the relative position for CC and WCC changes with network size.
Furthermore, CC produces the most non-singleton clusters for the large networks, but not for the medium networks.
These trends can be understood by realizing that both CM and WCC have the potential to take a non-singleton cluster from CC and break it up into subsets where some may be non-singletons, but possibly all are singletons. 
Thus, a given non-singleton cluster from the CC-treated SBM clustering can result in multiple non-singleton clusters or zero non-singleton clusters.

The trends we observe, therefore, indicate the following relative trends between the CC, WCC,  and CM treatments. CC and WCC treatments have similar effects on the cluster size distributions while CM treatment tends to produce larger and fewer clusters on a smaller subset of the nodes. However, CC does preserve a number of large clusters relative to WCC treatment.

\subsection*{Experiment 3: Impact of treatment on synthetic networks}
To evaluate the impact of our treatments on the clustering accuracy of SBM, we explored a range of LFR and RECCS synthetic networks, each modelled after clustered real-world networks. 
In Experiment 3a, we compare untreated and treated SBM clusterings to determine the best treatment, using the   four different accuracy metrics (NMI, ARI, AGRI, and RMI). 
In Experiment 3b, we then compared the best treatment for the selected SBM to  Leiden optimizing modularity or under the Constant Potts Model (CPM) for the same accuracy metrics.

\subsubsection*{Experiment 3a: Comparing the clustering accuracy of CC, WCC, and CM treatments of SBM  on synthetic networks}

\begin{figure}[!h]
\includegraphics[]{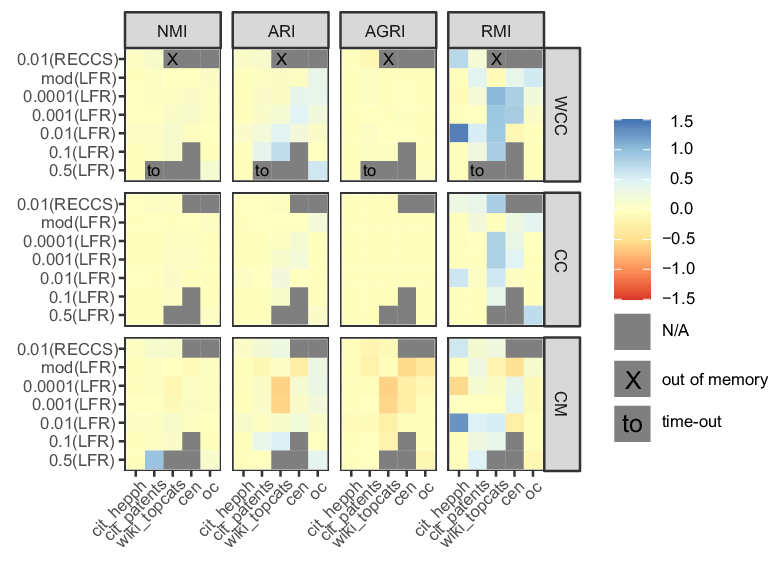}
\caption[Experiment 3a:  impact of treatment on NMI/ARI/AGRI/RMI scores of selected SBM on LFR and RECCS networks (heatmap)]{\textbf{Experiment 3a:  
Impact of treatment on NMI/ARI/AGRI/RMI scores of selected SBM on LFR and RECCS networks (heatmap)} WCC exceeds CC, and CM treatments in improving the clustering accuracy of SBM. Each subplot shows the results for one synthetic network (either LFR or RECCS), defined by the real-world network (vertical) axes and clustering (horizontal axes).
Gray boxes with ``to" indicate time-outs and those with ``X" indicate OOM (out-of-memory) errors.
Gray boxes without text (N/A in legend) are for networks that are not available; see text for explanation.
} \label{fig:3}
\end{figure}

We present a heatmap showing the impact of each treatment (CC, WCC, and CM) on the clustering accuracy for the four different measures in Fig~\ref{fig:3}.
Here, yellow indicates neutral impact, orange and red indicate that accuracy is reduced using the treatment, and blue indicates accuracy is improved using the treatment.
While the impact depends on the synthetic network and accuracy measure, we see that  WCC was in general neutral for NMI and AGRI, but neutral to beneficial to ARI and RMI.
CC showed similar trends as WCC but was not as beneficial for AGRI and RMI. 
Finally, CM was also mainly neutral for NMI, but for each of the other accuracy measures there are  synthetic networks where CM is clearly detrimental. 

Thus, overall WCC is the best treatment for the selected SBM, generally improving accuracy over the untreated version and over other treatments.
Moreover, while WCC always was neutral or beneficial for NMI and ARI accuracy,  there are 8 model conditions where using WCC reduced RMI or AGRI accuracy, usually by a small amount.
The largest reduction in accuracy is the reduction in RMI score from   92.4 to 83.1  on the LFR network for the CEN real-world network clustered using Leiden-CPM(0.01). 
 Nevertheless, despite these cases where WCC reduced RMI accuracy, the improvements in accuracy obtained using WCC were both larger and more frequent than the reductions in accuracy, as shown in the heatmap.
 For these reasons, we selected WCC as the preferred treatment.

Finally, WCC but not CC or CM had timeouts and memory errors.  On the RECCS network for wiki\_topcats clustered using Leiden-CPM(0.01),  WCC ran into a memory error with 256GB of RAM, and WCC ran out of time (the limit was 72 hours) when clustering the LFR network for the   cit\_patents network clustered using Leiden-CPM(0.5).  

\begin{figure}[!h]
\includegraphics[]{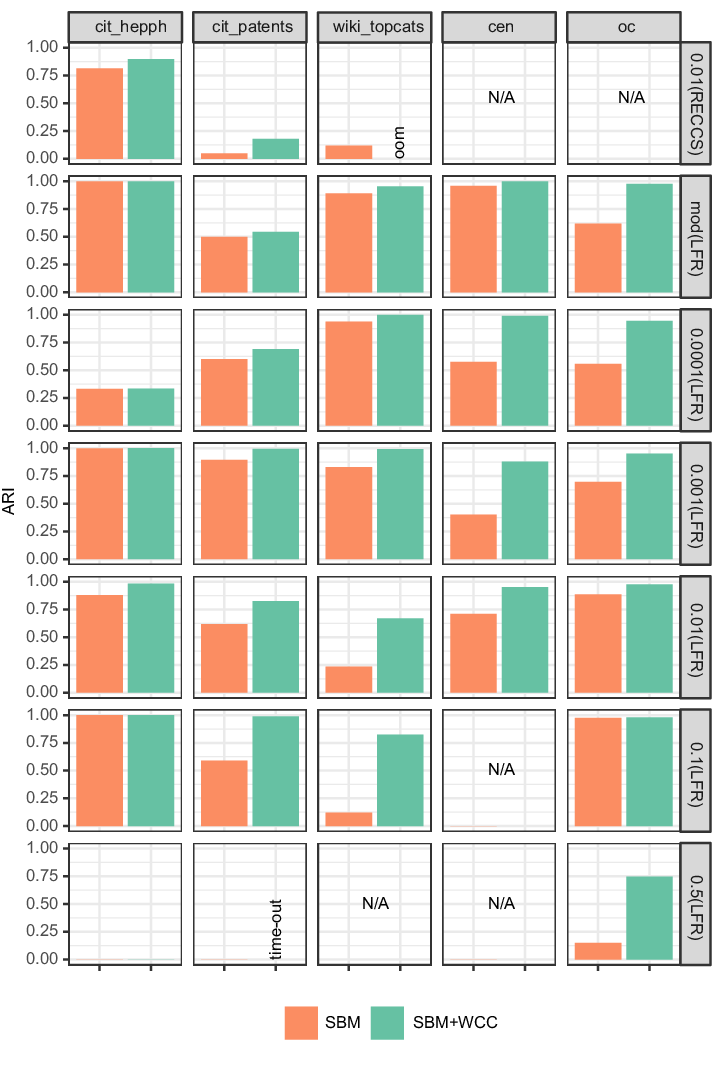}
\caption[Experiment 3a: Impact of WCC treatment on ARI scores of selected SBM  ]{\textbf{Experiment 3a: Impact of WCC treatment on ARI scores of selected SBM (bar chart).}  WCC treatment of SBM clusterings benefits accuracy. 
The subplots marked with ``N/A'' indicate networks that are not available; see text for explanation.
The subplot marked with ``time-out'' indicates that WCC failed to complete within 72 hours. The bottom left subplot appears empty due to
both SBM and SBM+WCC yielding   ARI accuracy of $0.0$.

}\label{fig:4}
\end{figure}

Given the trends so far, we select WCC as the preferred treatment for SBM clustering.
We then explore the impact of WCC on ARI clustering accuracy in depth in Fig~\ref{fig:4}, exploring the ARI accuracy scores for the untreated SBM and the WCC-treated SBM.

Note that for many of the cases where WCC has  neutral impact, the untreated SBM already has very high accuracy, leaving little room for improvement.   

\subsubsection*{Experiment 3b: Evaluation of clustering accuracy on synthetic networks}

In this section, we examine clustering accuracy for four different accuracy criteria for Leiden-CPM (resolution value 0.001) and Leiden-mod, each treated or untreated by CM and WCC, in comparison to SBM+WCC.

We consider the following two questions:
\begin{itemize}
    \item Question 1: How do the two Leiden methods respond to treatment?  Is CM better than WCC?
    \item Question 2: What is the best method in general for each accuracy criterion?
\end{itemize}

\begin{figure}[!h]
\includegraphics[]{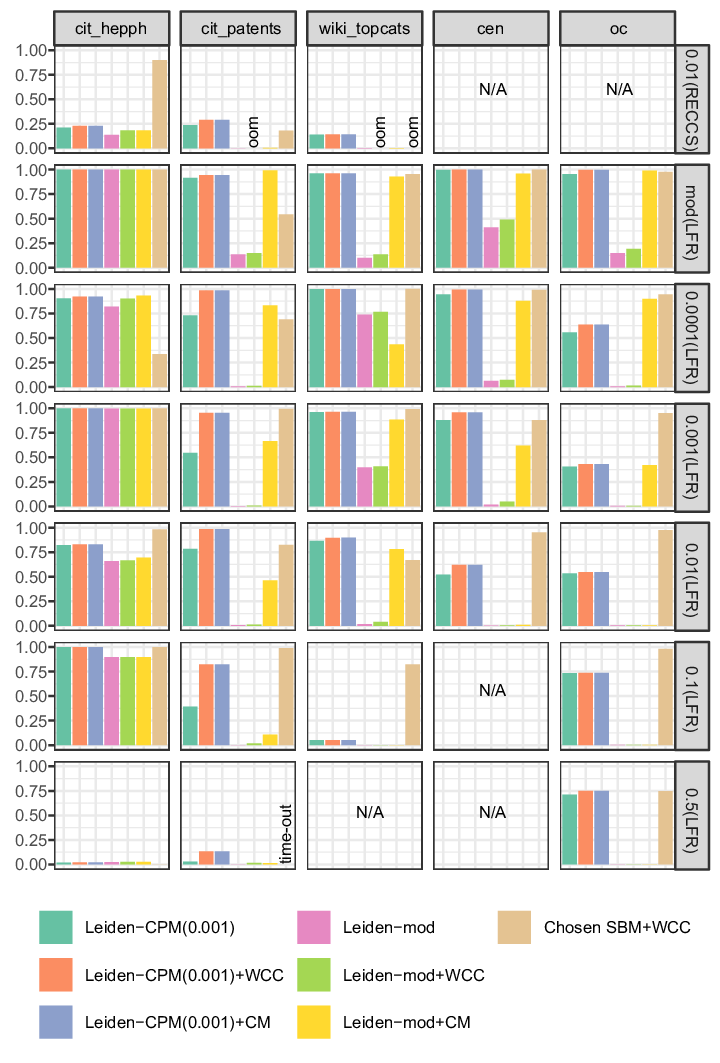}
\caption[Experiment 3b: ARI accuracies of SBM+WCC against various methods and their treatments on LFR and RECCS synthetic networks]{\textbf{Experiment 3b: ARI accuracies of SBM+WCC against various methods and their treatments on LFR and RECCS synthetic networks.} SBM+WCC is generally competitive with Leiden-CPM(0.001) and Leiden-mod. Each subplot shows results for one synthetic network (either LFR or RECCS), defined by the real-world network (vertical) axes and clustering (horizontal axes).
The ``N/A'' cells are those networks that are not available; see text for explanation.
There are  three out-of-memory (oom) entries (top row) and one ``time-out" entry (bottom row).
}
\label{fig:5}
\end{figure}

\begin{figure}[!h]
\includegraphics[]{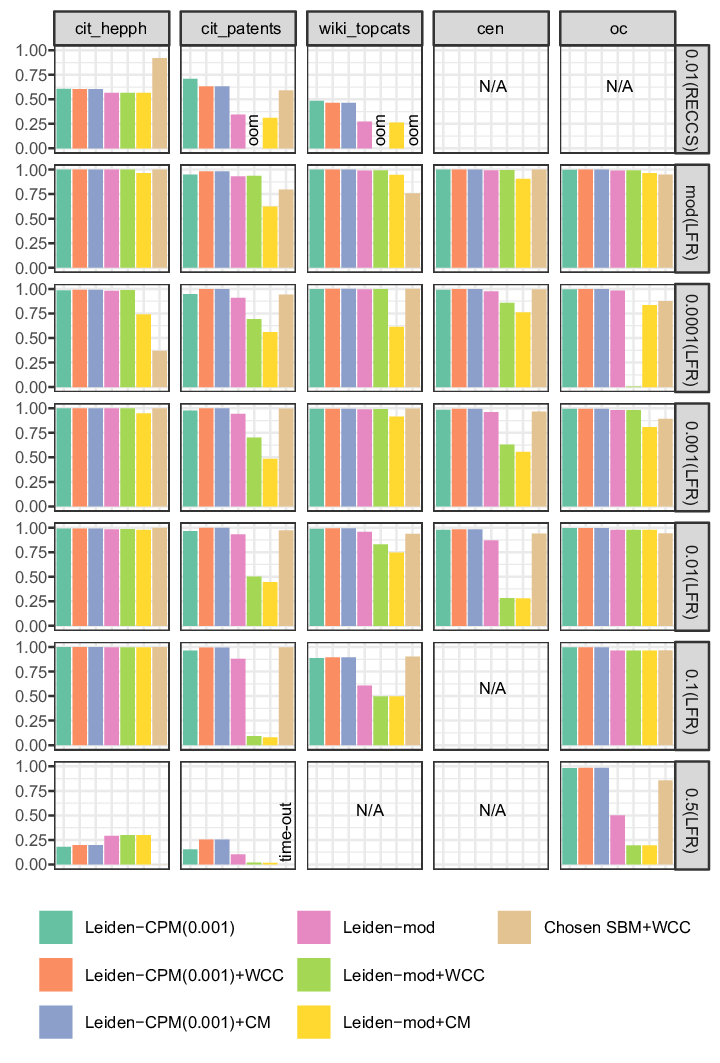}
\caption[Experiment 3b: AGRI accuracies of SBM+WCC against various methods and their treatments on LFR and RECCS synthetic networks]{\textbf{Experiment 3b: AGRI accuracies of SBM+WCC against various methods and their treatments on LFR and RECCS synthetic networks.} SBM+WCC is generally competitive with Leiden-CPM(0.001) and Leiden-mod.
Each subplot shows results for one synthetic network (either LFR or RECCS), defined by the real-world network (vertical) axes and clustering (horizontal axes).
The ``N/A'' cells are those networks that are not available; see text for explanation.
There are  three out-of-memory (oom) entries (top row) and one ``time-out" entry (bottom row).}
\label{fig:6}
\end{figure}

Accuracy results for ARI and AGRI are shown in 
Figs   \ref{fig:5} and \ref{fig:6}; 
and those for NMI and RMI are in S1 Appendix (Figs B and C).

We begin with Question 1.
For Leiden-CPM(0.001), the accuracy for the two treatments CM or WCC seem indistinguishable.
In addition, 
both are typically at least as good as  untreated Leiden-CPM(0.001) for NMI, ARI, and AGRI accuracy measures; for RMI there a few cases where CM and WCC reduce accuracy but more cases where they improve accuracy.
Overall, we can say CM and WCC have the same and generally positive impact for Leiden-CPM(0.001).
For Leiden-mod, CM and WCC have different impacts. For NMI and ARI, both tend to improve accuracy, but CM is better than WCC. 
For AGRI, they can hurt accuracy.
For RMI, they tend to be neutral or improve accuracy. 
Overall,  the WCC and CM treatments cannot be said to be generally beneficial for Leiden-mod.

For Question 2, it is clear that 
the top  methods are clearly Leiden-CPM(0.001)+WCC/CM and SBM+WCC, and each of these was much more reliable than any variant of Leiden-mod, i.e., treated or untreated.

The comparison between Leiden-CPM(0.001)+WCC/CM  and SBM+WCC depends very much on the accuracy measure and model condition, and neither really dominates the other. 
Nevertheless, we can make the following observations.
For NMI accuracy, the two methods were generally very close in accuracy and there were no cases where Leiden-CPM(0.001)+WCC/CM was much more accurate than SBM+WCC; however, there were some cases where SBM+WCC was much more accurate than Leiden-CPM(0.001)+WCC/CM (e.g., on the RECCS network for cit\_hepph). 
For ARI accuracy, there were cases where each method was substantially more accurate than the other, but the model conditions producing the biggest differences in accuracy favored SBM+WCC.  
For AGRI accuracy, while there were model conditions where SBM+WCC was somewhat more accurate than Leiden-CPM(0.001)+CM, in many model conditions Leiden-CPM(0.001)+CM had much better AGRI scores; hence, for this accuracy measure, Leiden-CPM(0.001)+CM is superior to SBM+WCC.
Finally, for RMI accuracy, the two methods are very close, but each method was much more accurate than the other for one model condition. 
Overall, SBM+WCC can be seen as generally equivalent for most accurate with Leiden-CPM(0.001)+CM in our experiments.

\clearpage
\subsection*{Experiment 4: Computational performance}

In general SBM+WCC was able to  complete on nearly every network we analyzed with the exception of one real-world network and two synthetic networks where SBM+WCC either a time-out or an out-of-memory error occurred. 

\begin{figure}[!h]
\includegraphics[]{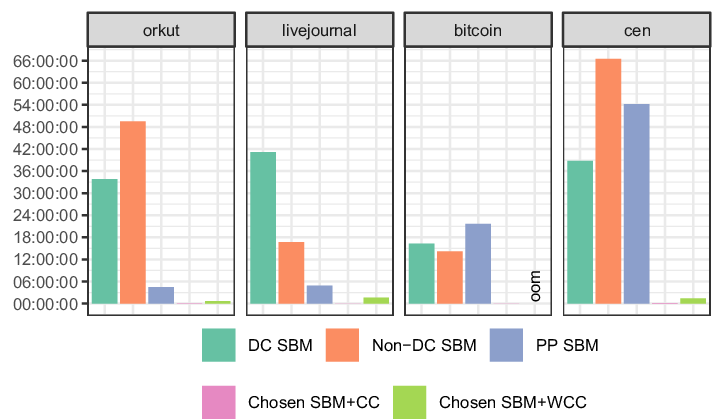}
\caption[Experiment 4: Runtime of SBM and SBM + treatments on large non-bipartite real-world networks]{\textbf{Experiment 4: Runtime of SBM and SBM + treatments on large non-bipartite real-world networks} The runtime of CC or WCC treatments on SBM clusterings is negligible compared to the runtime of SBM. Here we show the runtime of each method in hours:minutes:seconds format. For Chosen SBM+CC and Chosen SBM+WCC, only the time it took to run the treatment is shown. Number of nodes: orkut - 3,072,441; livejournal - 4,847,571; bitcoin - 6,336,770; cen - 13,989,436 %\textcolor{orange}{new figure}
} \label{fig:7}
\end{figure}

Here we describe the runtime on the four largest real-world networks from this study; see   Fig~\ref{fig:7}. 
The SBM+WCC pipeline involves computing an SBM clustering under each of the three models (degree corrected, non-degree corrected, and planted partition), and then following with WCC on the clustering that achieved the lowest description length.
As seen in Fig~\ref{fig:7}, by far the most computationally intensive part is computing the SBM clustering for each of the three models, which take between 5 and 66 hours each. 
Running  CC is negligible, completing in minutes, and running WCC finishes in under two hours on each network. For example, on the CEN, which has about 14 million nodes, the SBM model with the least run time was the degree corrected model, which took 38.7 hours. In comparison, WCC processing took 1.4 hours. Other models of SBM on the CEN were more expensive, with the non degree-corrected SBM at 66.5 hours and planted partition SBM at 54.2 hours.
Thus,  the time it took to cluster the networks using SBM far exceeded the time it took to process those clusterings through the CC and WCC treatments. 

\section*{Discussion} \label{sec:discussion}
There are two major trends observed in this study. First, that SBMs under all three tested models (degree corrected, non-degree corrected, and planted partition models) frequently produced clusters that are internally disconnected.
This tendency increased with the network size, but was evident even on relatively small networks.
Second, that WCC, one of the three post-processing techniques we examined,  trended towards an improvement in clustering accuracy, while CM had mixed impact, sometimes beneficial and sometimes detrimental. Here we attempt to explain why we see these trends.

\paragraph{Understanding why Degree Corrected SBM produces disconnected clusters } 

The observation that SBM clustering sometimes returns disconnected clusters has been made before \cite{peixoto2019bayesian}, using 
a network where each component is a clique, and the SBM is based on a very simple microcanonical model. 
That study also provided an explanation for this phenomenon by establishing that the description length for that model grows quadratically in the number of blocks (i.e., clusters), thus penalizing for a large number of clusters.

Here, we extend that observation and derivation to the case of the degree-corrected SBM.  We also examine description length scores for degree-corrected SBM clusterings on real-world networks and compare them to SBM+CC clusterings; this analysis reveals that a specific component of the description length dominates the score when we use the CC post-processing treatment.

Recall that for a given input network $N$, graph-tool under the degree-corrected model seeks a clustering (and hence an SBM) that minimizes the description length, 
making it an optimization problem. 

We now define the description length.  Given a proposed SBM, let

\begin{itemize}
    \item $A$ be the adjacency matrix, defined by the network $N$,
    \item $b$ be the block (cluster) assignment, which represents the clustering of $N$,
    \item $k$ be the degree vector, defined by $A$,  
    \item $e$ be the edge count matrix, defined  by $A$ and $b$.
\end{itemize}

Eq~(\ref{eq:dcsbm-dl}) provides the formula for the description length $\mathrm{DL}(A, b)$ of a network $A$ and a clustering $b$ under the Degree Corrected (DC) model:

\begin{eqnarray}
    \mathrm{DL}(A, b) = - \log p(A|b, e, k) - \log p(k|b, e) - \log p(b) - \log p(e)
\label{eq:dcsbm-dl}
\end{eqnarray}

Here, the description length can be decomposed into four parts, which are the negative logarithms of the model likelihood ($-\log p(A|b, e, k)$), the prior for the degree ($-\log p(k|b, e)$), the prior for the block assignment ($-\log p(b)$), and the prior for the edge count matrix ($-\log p(e)$). graph-tool provides the functionality for computing these quantities separately, which we will use for the following analyses.

\begin{figure}[!h]
    \centering
    \includegraphics[]{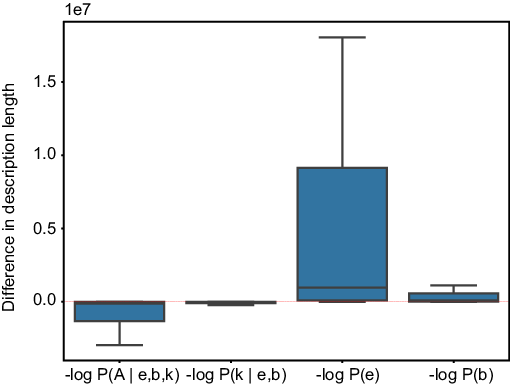}
    \caption[SBM(DC)+CC to SBM(DC) difference for description length components]{\textbf{SBM(DC)+CC to SBM(DC) difference for description length components.} We show the contribution of $-\log p(e)$ term on the description lengths of CC-treated SBM clusterings. Box plots of differences for components of the description length on the $71$ networks that select the SBM(DC) model for community detection and clusterings obtained by both SBM(DC) and SBM(DC)+CC for each network. The differences are SBM(DC)+CC - SBM(DC). Positive values indicate favoring not having the CC treatment. %\textcolor{red}{(The-Anh to Min) edit the figure if not already done}
    }
    \label{fig:8}
\end{figure}

We compute the components of the description length for all networks that select the SBM(DC) model, with and without the CC treatment; there are $71$ networks. Fig \ref{fig:8} illustrates the distribution of differences between SBM(DC)+CC (i.e., the output of the Degree Corrected model of SBM, but with CC treatment) and SBM(DC) (i.e., the output of the Degree Corrected model of SBM) for each component on all the networks. Since the difference is SBM(DC)+CC - SBM(DC), a positive difference means we do not favor the CC treatment. On all studied networks, the $-\log p(A|b, e, k)$ and $-\log p(k|b, e)$ component prefers connected clusters returned by the CC treatment. In contrast, the $-\log p(b)$ and $-\log p(e)$ components penalize the CC treatment.

We present a specific example of this phenomenon on the real-world network \texttt{linux}, detailed in Table \ref{tab:1}. The two components $-\log p(A|b, e, k)$ and $-\log p(k|b, e)$ of SBM(DC)+CC are lower, indicating an advantage of the CC treatment. However, the two components $-\log p(b)$ and $-\log p(e)$ of SBM(DC)+CC are higher, especially with the $- \log p(e)$ component having a much larger margin; this effectively negates all the advantage. As a result, with a lower description length, the untreated output of the Degree Corrected model of SBM is preferred over the same output with CC treatment. Had the $- \log p(e)$ {\em not been included} in the formula for the description length, SBM(DC)+CC would have been preferred for having a smaller description length.

\begin{table}[!ht]
\centering
\caption[Components of description lengths on \texttt{linux}.]{\textbf{Components of description lengths on \texttt{linux}.}}
\label{tab:1}
\begin{tabular}{lrrr}
\hline
Quantity & SBM(DC) & SBM(DC)+CC & Difference \\
\hline
$-\log p(A|b, e, k)$ & 699k & 316k & -383k \\
$-\log p(k|b, e)$ & 96k & 45k & -51k  \\
$-\log p(b)$ & 147k & 257k & 110k \\
$-\log p(e)$ & 51k & 1,585k & 1,534k \\
\hline
DL$(A, b)$ & 993k & 2,202k & 1,209k \\
\hline
\end{tabular}
\begin{flushleft}
The last row sums up the values in the previous four rows. The difference is SBM(DC)+CC - SBM(DC); so, a negative value means favoring SBM(DC) with CC treatment and a positive value means favoring untreated SBM(DC).
\end{flushleft}
\end{table}

We investigated all networks to see if the CC treatment is preferred when we remove the $-\log p(e)$ component. Our investigation shows, for $59$ networks out of $71$ networks, removing the $-\log p(e)$ component will result in a lower description length for the clustering output with CC treatment. Thus, the $-\log p(e)$ component accounts for $83.1\%$ of the cases where SBM(DC) without CC treatment is preferred over SBM(DC) with CC treatment.

We now examine  the formula for the $-\log p(e)$ component of SBM(DC). For undirected graphs, the formula is given by
\begin{eqnarray}
    - \log p(e) = \log \begin{pmatrix}
        B(B + 1)/2 + E - 1 \\
        E
    \end{pmatrix}
\end{eqnarray}
where $B$ is the number of blocks and $E$ is the number of edges. The formula shows that $-\log p(e)$ will increase as $B$ increases. 
Moreover, since $E>0$ is fixed,   $-\log p(e) = O(\log B)$.
Thus, this favors a small number $B$ of blocks.

Recall that both the CC and WCC treatments tend to increase the number of blocks $B$, and that even CC increases this number whenever any of the clusters are disconnected.
Since increasing the number of blocks makes  $-\log p(e)$ larger, this means that such treatments will tend to result in larger description lengths---which will not be favored by SBM.
Since the description length is impacted strongly by $-\log p(e)$, this
explains why SBM clustering will tend to prefer clusterings that have internally disconnected clusters rather than their CC-treated versions.

\paragraph{Understanding why SBM+WCC is more accurate than SBM+CM }
The second major observation is that WCC generally was neutral or improved SBM clustering accuracy, but CM had mixed impact, and even sometimes decreased accuracy.

We offer a hypothesis for this trend.
Note that if an SBM cluster is internally disconnected, then during the SBM+CM process, it is first split into its components, and then each of the components is re-clustered using SBM.
Our study suggests that the individual clusters in the new SBM clustering are likely to be internally disconnected.  When this happens, the CM process will continue to iterate and re-cluster yet again the individual components of the clusters.
This repeated iteration will break each of the original clusters into many smaller clusters, and reduce accuracy.

In contrast, the WCC process never re-clusters -- it only breaks clusters into well-connected subclusters.  Thus, it will first break a cluster that contains multiple components into its constituent components, and then remove small edge cuts until the cluster is well-connected. This process will of course iterate but the overall number of times it breaks a cluster into pieces is less than what CM does.

This hypothesis is consistent with the node coverage of SBM+CM being smaller than the node coverage of SBM+WCC on the medium and large networks, as we observed in this study (see Table A in S1 Appendix). 

In contrast, consider the impact of WCC and CM on Leiden-CPM or Leiden-mod clusterings. In those cases, using WCC and CM often had the same impact on accuracy, and using CM rarely reduced accuracy.
Note that Leiden never produces disconnected ground truth clusters, unlike SBM, but can produce poorly connected clusters.  
Thus, using CM will reduce node coverage more than using WCC, but the 
degree of the reduction  produced by CM on a Leiden-CPM or Leiden-mod clustering is less than that of CM on an SBM clustering.
This is perhaps why CM is generally beneficial for Leiden clusterings, and not as generally beneficial for SBM clusterings.

\section*{Conclusion} \label{sec:conclusion}

This study provides evidence that SBMs used for community detection can result in internally disconnected and poorly connected clusters. We  presented the Well-Connected Clusters (WCC) technique, which repeatedly removes small edge cuts until all clusters are well-connected.  We showed that WCC is fast enough to use on very large networks with millions of nodes, and that applying WCC to SBM clusterings can improve accuracy on synthetic networks.  Thus, WCC  is a simple method that addresses this limitation of SBMs.

Future work should examine other ways of improving connectivity of clusters produced by SBMs.
For example, modifying the search strategy to require that all valid clusterings produce connected clusters would address the most egregious problem of producing internally disconnected clusters.
Other approaches should be developed to improve  edge connectivity in the clusters produced by the SBM software.

\section*{Supporting information}
\textbf{Supplementary Materials Document}  This  PDF document contains additional details about the RECCS generation protocol and additional results provided in 3 supplementary tables and 3 supplementary figures.

\section*{Funding} \label{sec:funding}
This work was supported in part by the Illinois-Insper partnership. The authors thank the Illinois Computes Program for allocations of cluster computing time.

% \nolinenumbers
\bibliography{clustering}

\begin{thebibliography}{10}

\bibitem{kannan2004clusterings}
Kannan R, Vempala S, Vetta A.
\newblock On clusterings: Good, bad and spectral.
\newblock Journal of the ACM (JACM). 2004;51(3):497--515.

\bibitem{traag2019louvain}
Traag VA, Waltman L, Van~Eck NJ.
\newblock {From Louvain to Leiden: guaranteeing well-connected communities}.
\newblock Scientific Reports. 2019;9(1):1--12.

\bibitem{blondel2008fast}
Blondel VD, Guillaume JL, Lambiotte R, Lefebvre E.
\newblock Fast unfolding of communities in large networks.
\newblock Journal of statistical mechanics: theory and experiment. 2008;2008(10):P10008.

\bibitem{park2024improved-arxiv}
Park M, Feng DW, Digra S, Vu-Le TA, Chacko G, Warnow T. Improved Community Detection using Stochastic Block Models; 2024.
\newblock arXiv preprint arXiv:2408.10464.

\bibitem{graph-tool}
Peixoto TP.
\newblock The graph-tool python library.
\newblock figshare. 2014;doi:{10.6084/m9.figshare.1164194}.

\bibitem{park2024}
Park M, Tabatabaee Y, Ramavarapu V, Liu B, Pailodi VK, Ramachandran R, et~al.
\newblock Well-connectedness and community detection.
\newblock PLOS Complex Systems. 2024;1(3):e0000009.

\bibitem{leiden-code}
Traag V. Leiden Algorithm: leidenalg; 2019.
\newblock \url{https://github.com/vtraag/leidenalg}.

\bibitem{rosvall2009map}
Rosvall M, Axelsson D, Bergstrom CT.
\newblock The map equation.
\newblock The European Physical Journal Special Topics. 2009;178(1):13--23.

\bibitem{Wedell2022}
Wedell E, Park M, Korobskiy D, Warnow T, Chacko G.
\newblock Center-periphery structure in research communities.
\newblock Quantitative Science Studies. 2022;3(1):289--314.

\bibitem{VanDongen2008}
Dongen SV.
\newblock Graph clustering via a discrete uncoupling process.
\newblock {SIAM} Journal on Matrix Analysis and Applications. 2008;30(1):121--141.

\bibitem{lfr-generation-code}
Tabatabaee Y. Emulating real networks using {LFR} graphs; 2023.
\newblock \url{https://github.com/ytabatabaee/emulate-real-nets}.

\bibitem{hubert1985comparing}
Hubert L, Arabie P.
\newblock Comparing partitions.
\newblock Journal of classification. 1985;2:193--218.

\bibitem{anne2024synthetic}
Anne L, Vu-Le TA, Park M, Warnow T, Chacko G.
\newblock Synthetic Networks That Preserve Edge Connectivity.
\newblock arXiv preprint arXiv:240813647. 2024;.

\bibitem{peixoto2020netzschleuder}
Peixoto TP. The {Netzschleuder} network catalogue and repository; 2020.
\newblock \url{https://networks.skewed.de}.

\bibitem{lancichinetti2008benchmark}
Lancichinetti A, Fortunato S, Radicchi F.
\newblock Benchmark graphs for testing community detection algorithms.
\newblock {P}hysical {R}eview E. 2008;78(4):046110.

\bibitem{illinoisdatabankIDB-9805305}
Anne L, Park M, Warnow T, Chacko G. Synthetic Networks For Benchmarking; 2025.
\newblock Available from: \url{https://doi.org/10.13012/B2IDB-9805305_V1}.

\bibitem{dcsbm}
Karrer B, Newman ME.
\newblock Stochastic blockmodels and community structure in networks.
\newblock Physical Review E—Statistical, Nonlinear, and Soft Matter Physics. 2011;83(1):016107.

\bibitem{holland1983sbm}
Holland PW, Laskey KB, Leinhardt S.
\newblock Stochastic blockmodels: First steps.
\newblock Social Networks. 1983;5(2):109--137.

\bibitem{ppsbm}
Zhang L, Peixoto TP.
\newblock Statistical inference of assortative community structures.
\newblock Physical Review Research. 2020;2(4):043271.

\bibitem{henzinger2018practical}
Henzinger M, Noe A, Schulz C, Strash D.
\newblock Practical Minimum Cut Algorithms.
\newblock {ACM} Journal of Experimental Algorithmics. 2018;23:1--22.

\bibitem{thomas2006elements}
Thomas M, Joy AT.
\newblock Elements of Information Theory.
\newblock Wiley-Interscience; 2006.

\bibitem{poulin2020comparing}
Poulin V, Th{\'e}berge F.
\newblock Comparing graph clusterings: Set partition measures vs. graph-aware measures.
\newblock IEEE Transactions on Pattern Analysis and Machine Intelligence. 2020;43(6):2127--2132.

\bibitem{newman2020improved}
Newman ME, Cantwell GT, Young JG.
\newblock Improved mutual information measure for clustering, classification, and community detection.
\newblock Physical Review E. 2020;101(4):042304.

\bibitem{peixoto2019bayesian}
Peixoto TP.
\newblock 11.
\newblock In: Bayesian Stochastic Blockmodeling. John Wiley \& Sons, Ltd; 2019. p. 289--332.
\newblock Available from: \url{https://onlinelibrary.wiley.com/doi/abs/10.1002/9781119483298.ch11}.

\end{thebibliography}
% Either type in your references using
% \begin{thebibliography}{}
% \bibitem{}
% Text
% \end{thebibliography}
%
% or
%
% Compile your BiBTeX database using our plos2015.bst
% style file and paste the contents of your .bbl file
% here. See http://journals.plos.org/plosone/s/latex for 
% step-by-step instructions.
% 
% \begin{thebibliography}{10}

% \bibitem{bib1}
% Conant GC, Wolfe KH.
% \newblock {{T}urning a hobby into a job: how duplicated genes find new
%   functions}.
% \newblock Nat Rev Genet. 2008 Dec;9(12):938--950.

% \bibitem{bib2}
% Ohno S.
% \newblock Evolution by gene duplication.
% \newblock London: George Alien \& Unwin Ltd. Berlin, Heidelberg and New York:
%   Springer-Verlag.; 1970.

% \bibitem{bib3}
% Magwire MM, Bayer F, Webster CL, Cao C, Jiggins FM.
% \newblock {{S}uccessive increases in the resistance of {D}rosophila to viral
%   infection through a transposon insertion followed by a {D}uplication}.
% \newblock PLoS Genet. 2011 Oct;7(10):e1002337.

% \end{thebibliography}

\end{document}

%\mainmatter              % start of a contribution
%
\title{Supplementary Materials for Improved Community Detection using Stochastic Block Models}
%
%\titlerunning{Running title}  % abbreviated title (for running head)
%                                     also used for the TOC unless
%                                     \toctitle is used
%

\author{Minhyuk Park\textsuperscript{1} \and 
Daniel Wang Feng\textsuperscript{1} \and
Siya Digra\textsuperscript{1} \and 
The-Anh Vu-Le\textsuperscript{1} \and 
Lahari Anne \textsuperscript{1} \and
George Chacko \textsuperscript{1} \and 
Tandy Warnow\textsuperscript{1}}

\date{}
% \authorrunning{Park et al.} % abbreviated author list (for running head)

% \institute{Department of Computer Science, University of Illinois Urbana-Champaign, Urbana IL 61801\\
% \email{\{minhyuk2,warnow,chackoge\}@illinois.edu}\\ 
% }

\maketitle % typeset the title of the contribution

\textsuperscript{1}\emph{Siebel School of Computing and Data Science, University of Illinois Urbana-Champaign, Urbana, IL 61801. \{minhyuk2, chackoge, warnow\}@illinois.edu}\}

% \clearpage
\tableofcontents
\listoffigures
\listoftables

\clearpage
\section{Additional Figures}
\begin{figure}[!ht]
\centering
\includegraphics[]{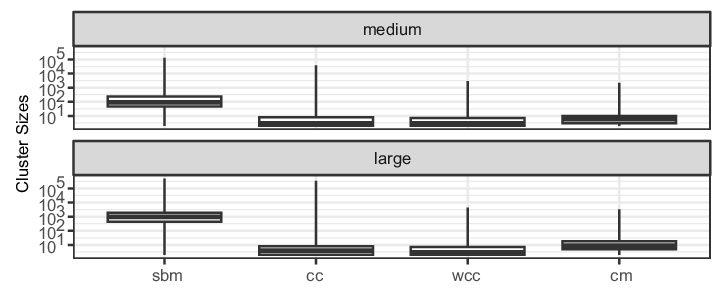}
\caption[Experiment 2: Impact of treatment on the non-singleton cluster size distributions of SBM clusterings on medium and large real-world networks (non-bipartite)]{\textbf{Experiment 2: Impact of treatment on the non-singleton cluster size distributions of SBM clusterings on medium and large real-world networks (non-bipartite) }
We show that SBM+WCC is competitive compared to Leiden-CPM(0.001) or Leiden-mod with the exception of very few cases. For 84 real-world networks grouped by the number of nodes, labeled medium ($1000 < n < 1,000,000$) and large ($n \geq{} 1,000,000$), we show the distribution of non-singleton cluster sizes resulting from clustering these networks with the lowest description model of SBM as well as its CC, WCC, and CM treatment clusterings where SBM has the largest cluster sizes followed by CM and then CC and WCC.. The boxplots are shown with a log scale y-axis with the whiskers indicating the smallest and largest cluster sizes for each clustering method in a dataset group. On one of the lage-sized datasets, WCC encountered an out-of-memory error on a machine with 256GB of RAM and is not included in any of the statistics for SBM, SBM+CC, or SBM+CM. The median and maximum cluster sizes for each box plot are listed here. Medium group median/max: SBM: 91/132109, SBM+CC: 3/38539, SBM+WCC: 3/2966, SBM+CM: 6/2169. Large group median/max: SBM: 933/262816, SBM+CC: 4/10187, SBM+WCC: 3/4387, SBM+CM: 9/3258% }\textbf{\textcolor{red}{Re-make table without bipartite graphs today (1/17) - update: it's been updated for both the actual figure and the caption listing the numbers. The boxes shifted down.}

% The distribution of non-singleton cluster sizes  is shown as a boxplot for the selected SBM and its treatments. The y-axis is plotted on a log scale with the whiskers indicating the minimum and maximum cluster sizes in all of the networks in the group.
% Both groups and treatments have minimum cluster size of 2 for SBM clusterings whether treated or not, but differ in the medians and maxes, as follows.
}\label{fig:s1}
\end{figure}
\begin{figure}[!h]
\centering
\includegraphics[]{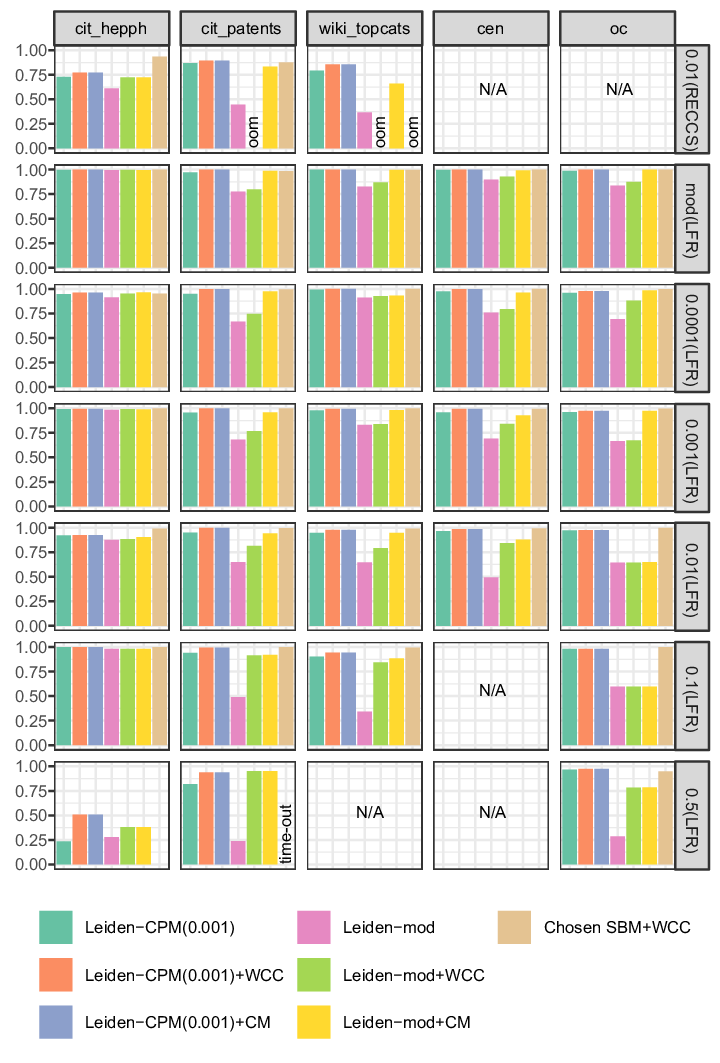}
\caption[Experiment 3b: NMI accuracies of SBM+WCC against various methods and their treatments on LFR and RECCS synthetic networks]{\textbf{Experiment 3b: NMI accuracies of SBM+WCC against various methods and their treatments on LFR and RECCS synthetic networks.} We show that SBM+WCC is competitive compared to Leiden-CPM(0.001) or Leiden-mod with the exception of very few cases. Each subplot shows results for one synthetic network (either LFR or RECCS), defined by the real-world network (vertical) axes and clustering (horizontal axes).
The ``N/A'' cells are those networks that are not available; see text for explanation.
There are  three out-of-memory (oom) entries (top row) and one ``time-out" entry (bottom row).}
\label{fig:s2}
\end{figure}

\begin{figure}[!h]
\centering
\includegraphics[]{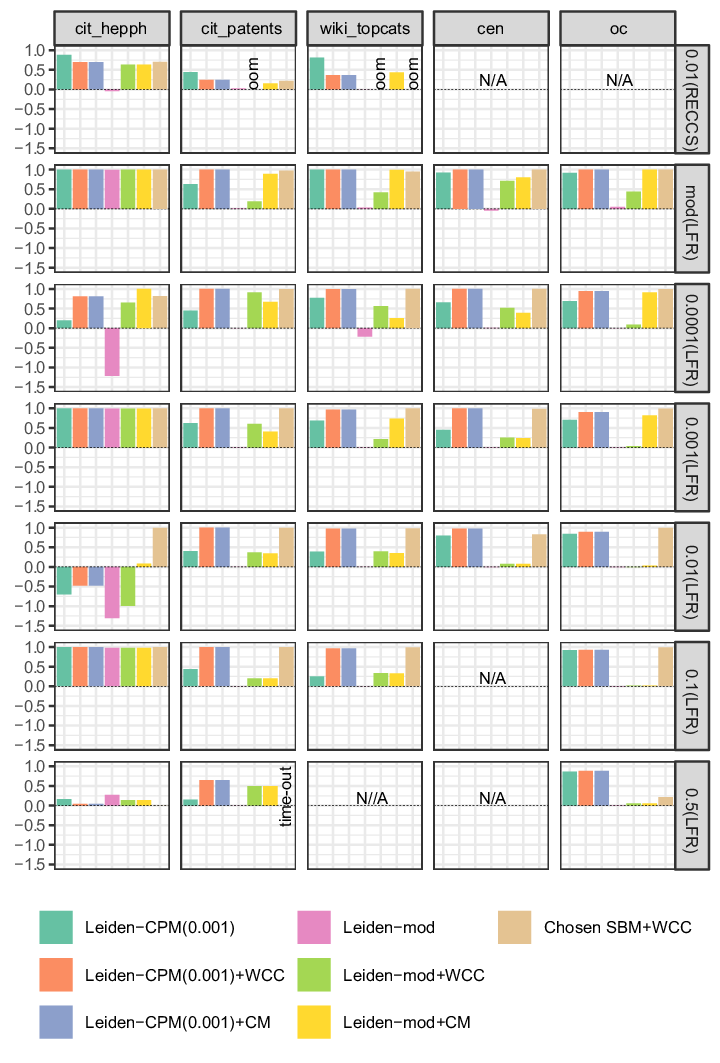}
\caption[Experiment 3b: RMI accuracies of SBM+WCC against various methods and their treatments on LFR and RECCS synthetic networks]{\textbf{Experiment 3b: RMI accuracies of SBM+WCC against various methods and their treatments on LFR and RECCS synthetic networks.}
Each subplot shows results for one synthetic network (either LFR or RECCS), defined by the real-world network (vertical) axes and clustering (horizontal axes).
The ``N/A'' cells are those networks that are not available; see text for explanation.
There are  three out-of-memory (oom) entries (top row) and one ``time-out" entry (bottom row).}
\label{fig:s3}
\end{figure}
\clearpage
\section{Extra Tables}
\begin{table}[!ht]
\centering
\caption[Experiment 2: Impact of treatment on node coverage on non-bipartite real-world networks]{\textbf{Experiment 2: impact of treatment on node coverage on non-bipartite real-world networks. } For 84 real-world networks (non-bipartite) grouped by the number of nodes, labeled small ($n \leq{} 1000$), medium ($1000 < n < 1,000,000$), and large ($n \geq{} 1,000,000$), we show the percentage of nodes in non-singleton clusters for the selected SBM clusterings (i.e., model achieving the lowest description length), with and without CC, WCC, and CM treatments. The node coverages are averaged across networks. On one of the lage-sized datasets, WCC encountered an out-of-memory error on a machine with 256GB of RAM and is not included in any of the statistics for SBM, SBM+CC, or SBM+CM.}
\label{tab:s1}
\begin{tabular}{lccc}
\hline
clustering & \multicolumn{3}{c}{node coverage} \\
& small  & medium  & large \\
\hline
Selected SBM & 74\% & 100\% & 100\% \\
Selected SBM + CC& 74\% & 64\% & 54\% \\ % used to be 59
Selected SBM + WCC& 67\% & 48\% & 38\% \\
Selected SBM + CM& 67\% & 35\% & 32\% \\ % used to be 32
\hline
\end{tabular}
\begin{flushleft}

\end{flushleft}
\end{table}

\begin{table}[!ht]
\centering
\caption[Experiment 2: Impact of treatment on node coverage on bipartite real-world networks]{\textbf{Experiment 2: impact of treatment on node coverage on bipartite real-world networks. } For 35 real-world networks (bipartite) grouped by the number of nodes, labeled small ($n \leq{} 1000$), medium ($1000 < n < 1,000,000$), and large ($n \geq{} 1,000,000$), we show the percentage of nodes in non-singleton clusters for the selected SBM clusterings (i.e., model achieving the lowest description length), with and without CC, WCC, and CM treatments. The node coverages are averaged across networks.}
\label{tab:s2}
\begin{tabular}{lccc}
\hline
clustering & \multicolumn{3}{c}{node coverage} \\
& small  & medium  & large \\
\hline
Selected SBM & 100\% & 100\% & 100\% \\
Selected SBM + CC& 100\% & 11\% & 27\% \\
Selected SBM + WCC& 82\% & 9\% & 12\% \\
Selected SBM + CM& 82\% & 0\% & 5\%\\
\hline
\end{tabular}
\begin{flushleft}

\end{flushleft}
\end{table}

\begin{table}[!ht]
% \begin{adjustwidth}{-2.25in}{0in} % Comment out/remove adjustwidth environment if table fits in text column.
\centering
\caption[Experiment 2: Average number of non-singleton clusters for the selected SBM, both treated and untreated, per non-bipartite real-world network group]{\textbf{Experiment 2: Average number of non-singleton clusters for the selected SBM, both treated and untreated, per non-bipartite real-world network group} For 84 real-world networks grouped by the number of nodes, labeled small ($n \leq{} 1000$),  medium ($1000 < n < 1,000,000$), and large ($n \geq{} 1,000,000$), we show the average number of non-singleton clusters for each method. The model of SBM for each dataset is determined by whichever model, chosen from degree corrected, non degree corrected, and planted partition, produced the clustering with the lowest description length. The number of clusters are averaged across networks. The follow-up CC, WCC, and CM treatments were done on the lowest description length SBM clustering. On one of the lage-sized datasets, WCC encountered an out-of-memory error on a machine with 256GB of RAM and is not included in any of the statistics for SBM, SBM+CC, or SBM+CM. }
\begin{tabular}{@{}lccc@{}}
\hline
 & small & medium & large \\
 \hline
Selected SBM & 1.43& 398.96& 3024.50\\ % used to be 2539.80
Selected SBM + CC & 1.43& 3214.05& 25955.00\\ % used to be 66431.20
Selected SBM + WCC & 1.43& 4171.45& 45969.25\\ 
Selected SBM + CM & 1.43& 2426.70 & 48331.75 \\ % used to be 39814.00
\hline
\end{tabular}
\begin{flushleft}

\end{flushleft}
\label{tab:s3}
% \end{adjustwidth}
\end{table}
\clearpage
\section{Software} \label{sec:software}
The main implementation for WCC, and CC, are written in C++ and is hosted at this repository \url{https://github.com/MinhyukPark/constrained-clustering}.

%\section*{Commands and Versions} \label{sec:commands-and-versions}
 CC and WCC treatments on Leiden-\{mod, cpm\} on LFR networks were done with the Python code base. CC and WCC treatments on all models of SBM were done with the C++ code base. CC of both code bases are deterministic while WCC of each code bases may result in different outcomes each time depending on which mincut is selected when there is a tie. CM was only done through the Python code base.

\paragraph{SBM} graph-tool can be found at \url{https://graph-tool.skewed.de/static/doc/} \cite{graph-tool}. We used version 2.59 on Python 3.9.18.
\begin{lstlisting}[language=Python]
import graph_tool.all as gt
g = gt.load_graph_from_csv(inputGraphName, csv_options = {'delimiter': '\t'})
```
clustering = gt.minimize_blockmodel_dl(g, state=gt.BlockState, state_args=dict(deg_corr=True)) # for DC sbm
clustering = gt.minimize_blockmodel_dl(g, state=gt.BlockState, state_args=dict(deg_corr=False)) # for Non DC sbm
clustering = gt.minimize_blockmodel_dl(g, state=gt.PPBlockState) # for PP sbm
```
description_length = clustering.entropy()
membershipFile = open(outputMemberName, "w+")
blockMembership = clustering.get_blocks()
for v in g.vertices():
    nodeId = g.vp.name[v]
    cur_block = blockMembership[v]
    if cur_block < 0 or cur_block > num_nodes_total - 1:
        continue
    membershipFile.write((str)(nodeId) + "\t" + (str)(cur_block) + "\n")
membershipFile.close()
\end{lstlisting}

\paragraph{CC} CC code can be found at \url{https://github.com/MinhyukPark/constrained-clustering}. The tag is v1.1.0.
\begin{lstlisting}
./constrained_clustering MincutOnly --edgelist <tab separated edgelist network> --existing-clustering <input clustering> --num-processors <maximum allowed parallelism> --output-file <output file path> --log-file <log file path> --log-level 1 --connectedness-criterion 0
\end{lstlisting}

\paragraph{WCC} WCC code can be found at \url{https://github.com/MinhyukPark/constrained-clustering} The tag is v1.1.0.
\begin{lstlisting}
./constrained_clustering MincutOnly --edgelist <tab separated edgelist network> --existing-clustering <input clustering> --num-processors <maximum allowed parallelism> --output-file <output file path> --log-file <log file path> --log-level 1 --connectedness-criterion 1
\end{lstlisting}

\paragraph{CM} CM code can be found at \url{https://github.com/illinois-or-research-analytics/cm_pipeline} The commit was a1c1d29.
\begin{lstlisting}
python3 -m hm01.cm -i <tab separated edgelist network> -e <input clustering> -o <output file path> -c external -cfile <clusterer file path e.g., path to hm01/clusterers/external_clusterers/sbm_wrapper.py> --threshold <threhsold e.g., 1log10> -cargs <clusterer args path (json detail found in repository) > --nprocs <number of processors>
\end{lstlisting}
\clearpage
\section{Real-world networks}
The lowest description length SBM clustering on ceo\_club and elite networks from the small networks returned all singleton clusters. These datasets were excluded from our study.\\

\noindent{}\textbf{Small networks (10) from \cite{peixoto2020netzschleuder} sorted by increasing node count} \\

\noindent{}\textit{bipartite}: sa\_companies; ceo\_club; elite; \\

\noindent{}\textit{non-bipartite}: november17; dutch\_criticism; macaque\_neural; sp\_kenyan\_households; contiguous\_usa; cs\_department; dolphins \\
% The selected SBM clustering of the ceo\_club and elite networks returned only singleton clusters.

\noindent{}\textbf{Medium-size networks (103) from \cite{peixoto2020netzschleuder} sorted by increasing node count} \\

\noindent{}\textit{bipartite}: plant\_pol\_robertson; escorts; movielens\_100k; nematode\_mammal; paris\_transportation; jester; dbpedia\_writer; digg\_votes; dbtropes\_feature; dbpedia\_starring; github; dbpedia\_recordlabel; dbpedia\_producer; dbpedia\_location; dbpedia\_occupation; dbpedia\_genre; discogs\_label; wiki\_article\_words; corporate\_directors; lkml\_thread; bookcrossing; flickr\_groups; visualizeus; dbpedia\_country; stackoverflow; eu\_procurements; epinions; citeulike; dbpedia\_team; bibsonomy \\

\noindent{}\textit{non-bipartite}: dnc; uni\_email; polblogs; faa\_routes; netscience; new\_zealand\_collab; collins\_yeast; interactome\_stelzl; bible\_nouns; at\_migrations; interactome\_figeys; us\_air\_traffic; drosophila\_flybi; fly\_larva; interactome\_vidal; openflights; bitcoin\_alpha; fediverse; power; advogato; bitcoin\_trust; jung; reactome; jdk; elec; chess; sp\_infectious; wiki\_rfa; dblp\_cite; anybeat; chicago\_road; foldoc; inploid; google; marvel\_universe; fly\_hemibrain; internet\_as; word\_assoc; cora; lkml\_reply; linux; topology; email\_enron; pgp\_strong; facebook\_wall; slashdot\_threads; python\_dependency; marker\_cafe; epinions\_trust; slashdot\_zoo; twitter\_15m; prosper; wiki\_link\_dyn; livemocha; wikiconflict; lastfm\_aminer; wiki\_users; wordnet; douban; academia\_edu; google\_plus; libimseti; email\_eu; stanford\_web; dblp\_coauthor\_snap; notre\_dame\_web; citeseer; twitter; petster; yahoo\_ads; berkstan\_web; myspace\_aminer; google\_web \\

\noindent{}\textbf{Large networks (7) from \cite{peixoto2020netzschleuder} sorted by increasing node count} \\

\noindent{}\textit{bipartite}: reuters; discogs\_affiliation; amazon\_ratings; dblp\_author\_paper \\

\noindent{}\textit{non-bipartite}: hyves; livejournal; bitcoin \\

% WCC on the bitcoin network ran into memory errors with 256GB.
\noindent{}\textbf{Other datasets (2) from \cite{park2024}} \\

\noindent{}\textit{non-bipartite}: orkut; CEN (Curated Exosome Network) \\

\clearpage
\section{RECCS synthetic network generation protocol}
RECCS synthetic network generator takes two inputs: a network and clustering on that network. Using these two inputs, it can then output a synthetic network using the input clustering as ground-truth. For our study, we used three real-world networks (cit\_hepph, cit\_patents, and wiki\_topcats) and clustered each of them using Leiden-CPM(0.01). These real-world network and clustering pairs were given to RECCS to generate the final synthetic networks. The details for how to run RECCS and the software can be found at \url{https://github.com/illinois-or-research-analytics/lanne2_networks/tree/main/generate_synthetic_networks}.

\bibliography{clustering}